\documentclass[journal]{vgtc}                     

\onlineid{1531}

\vgtccategory{Research}

\vgtcpapertype{please specify}

\title{
OFERA: Blendshape-driven 3D Gaussian Control for\\ Occluded Facial Expression to Realistic Avatars in VR
}

\author{%
  \authororcid{Seokhwan Yang}{0009-0000-1097-646X},
  \authororcid{Boram Yoon}{0000-0003-3696-0145},
  \authororcid{Seoyoung Kang}{0000-0002-6143-4369},
  \authororcid{Hail Song}{0009-0006-4008-196X},
  \authororcid{Woontack Woo}{0000-0002-5501-4421}
}

\authorfooter{
  \item
  	Seokhwan Yang, Seoyoung Kang and Hail Song are with KAIST UVR Lab.
    E-mail: \{ysshwan147\,$|$\,sy1009kang\,$|$\,hail96\}@kaist.ac.kr\,.
  \item
  	Boram Yoon is with KAIST KI-ITC ARRC.
  	E-mail: boram.yoon1206@kaist.ac.kr\,.
  \item 
    Woontack Woo is with KAIST UVR Lab and KAIST KI-ITC ARRC. Corresponding Author.
  	E-mail: wwoo@kaist.ac.kr\,.
}

\abstract{%
We propose \textbf{OFERA}, a novel framework for real-time expression control of photorealistic Gaussian head avatars for VR headset users. Existing approaches attempt to recover occluded facial expressions using additional sensors or internal cameras, but sensor-based methods increase device weight and discomfort, while camera-based methods raise privacy concerns and suffer from limited access to raw data. 
To overcome these limitations, we leverage the blendshape signals provided by commercial VR headsets as expression inputs. Our framework consists of three key components: (1) Blendshape Distribution Alignment (BDA), which applies linear regression to align the headset-provided blendshape distribution to a canonical input space; (2) an Expression Parameter Mapper (EPM) that maps the aligned blendshape signals into an expression parameter space for controlling Gaussian head avatars; and (3) a Mapper-integrated Avatar (MiA) that incorporates EPM into the avatar learning process to ensure distributional consistency.
Furthermore, OFERA establishes an end-to-end pipeline that senses and maps expressions, updates Gaussian avatars, and renders them in real-time within VR environments.
We show that EPM outperforms existing mapping methods on quantitative metrics, and we demonstrate through a user study that the full OFERA framework enhances expression fidelity while preserving avatar realism. By enabling real-time and photorealistic avatar expression control, OFERA significantly improves telepresence in VR communication.
A project page is available at https://ysshwan147.github.io/projects/ofera/.
}


\teaser{
  \centering
  \includegraphics[width=0.9\linewidth]{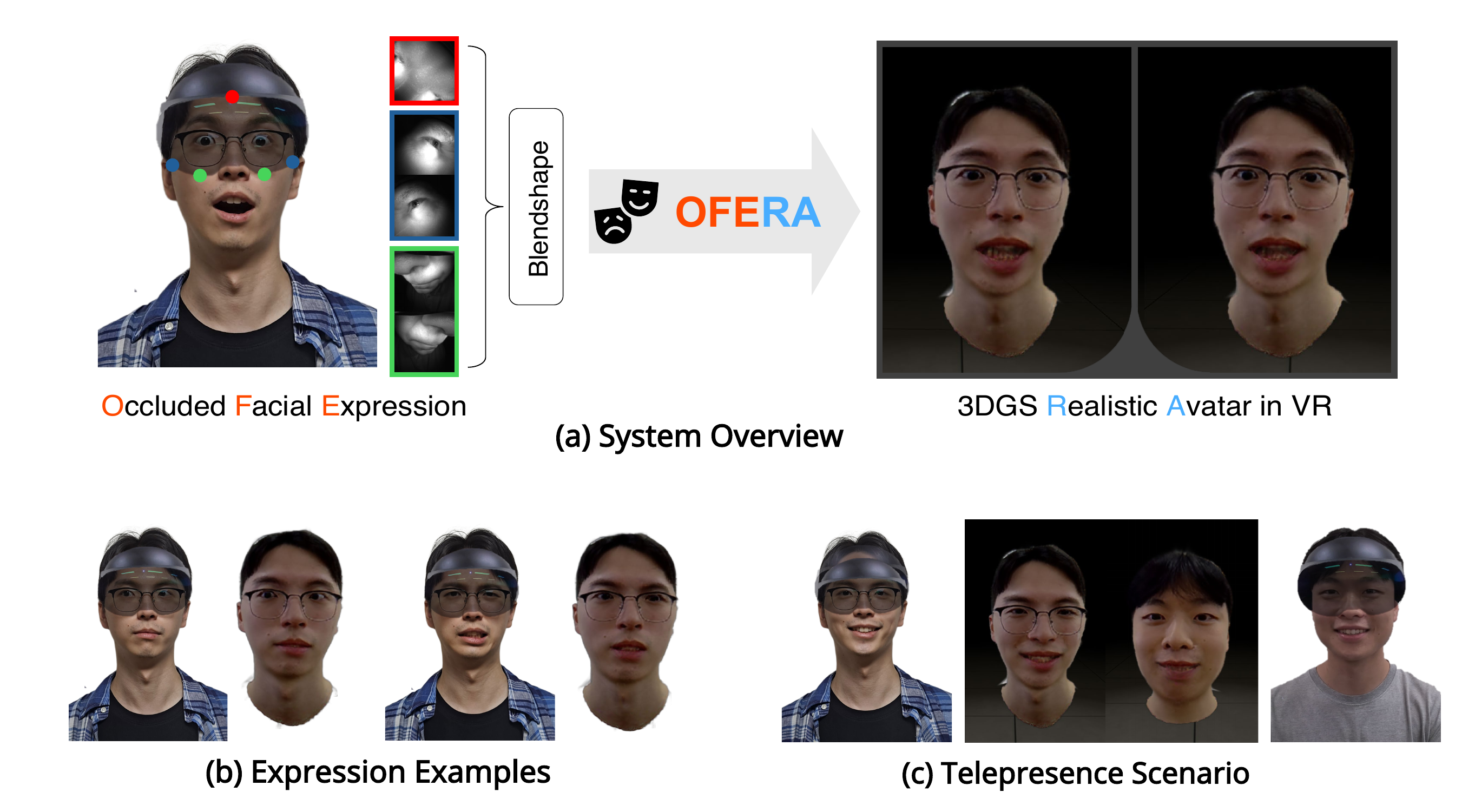}
  \caption{%
  We introduce \textbf{OFERA}, a system that transfers the facial expressions of VR headset users to photorealistic 3D Gaussian Splatting (3DGS)\cite{kerbl20233d} avatars using blendshape data. (a) Blendshape coefficients obtained from a VR headset are mapped by OFERA to animate a 3DGS avatar. (b) OFERA faithfully reproduces diverse facial expressions of VR headset users. (c) The system enables multi-user telepresence in VR, where users interact through realistic 3DGS avatars that convey their expressions. \textit{Note: The IR images shown in (a) are included for visualization only and sampled from the Ava-256\cite{martinez2024codec} dataset, since raw images from built-in headset cameras are not accessible. Images of users wearing headsets are created by overlaying semi-transparent headset renders on the original face images to illustrate headset-induced facial occlusion.}%
  }
  \label{fig:teaser}
}

\graphicspath{{figs/}{figures/}{pictures/}{images/}{./}} 

\usepackage[normalem]{ulem}

\usepackage{tabu}                      
\usepackage{booktabs}                  
\usepackage{lipsum}                    
\usepackage{mwe}                       
\usepackage{amssymb}
\usepackage{mathptmx}                  
\usepackage{amsmath}

\usepackage{colortbl}
\usepackage[table,xcdraw]{xcolor}

\newcommand{\removed}[1]{{\color{red}#1}}

\begin{document}



\maketitle

\section{Introduction}

In virtual reality (VR) communication, avatars that faithfully reproduce users' appearance and facial expressions are essential for enhancing the sense of telepresence. Two factors are particularly decisive: (i) how realistically avatars resemble users' facial appearance, and (ii) how accurately and responsively they reflect users' facial expressions in real-time. Prior studies have shown that the realism of avatar appearance and the responsiveness of expressions strongly influence social presence and collaboration outcomes in VR and mixed reality environments\cite{bailenson2004transformed,bailenson2006effect,baker2021avatar,combe2024exploring,lee2023effects,slater2010first}. These findings highlight that both visual fidelity and expressive capability are crucial for effective avatar-mediated communication.

Early approaches to avatar reconstruction relied on mesh-based parametric models that represent identity and expression with low-dimensional parameters\cite{blanz2023morphable,cao2013facewarehouse,li2017learning}. These models enabled efficient animation and controllable expression synthesis, and later works improved them with disentangled representations of pose and expression as well as finer geometry and appearance recovery\cite{deng2019accurate,feng2021learning,zhu2023facescape}. While lightweight and interpretable, mesh-based avatars remain visually limited compared to modern rendering techniques.

Neural rendering has since brought a major leap in realism. Neural Radiance Fields (NeRF)\cite{mildenhall2021nerf} first enabled photorealistic novel-view synthesis, and subsequent studies extended this framework to dynamic and controllable head avatars conditioned on expressions or audio signals\cite{gafni2021dynamic,grassal2022neural,guo2021ad}. Despite their realism, NeRF-based avatars suffer from long training and rendering times, limiting their use in real-time VR communication.

Recently, 3D Gaussian Splatting (3DGS)\cite{kerbl20233d} emerged as an efficient alternative, representing a scene as a collection of Gaussian primitives. This formulation supports interactive frame rates without sacrificing photorealistic detail. Building on this, recent work has demonstrated animatable 3DGS head avatars with improved expression control, relightability, and cross-subject generalization\cite{qian2024gaussianavatars,wang2025gaussianhead,zhang2025fate}. These advances make 3DGS particularly well-suited for VR communication. Nevertheless, most pipelines still assume unobstructed facial images, which is infeasible when VR headsets occlude large parts of users' faces.

To address this issue, prior works have explored two directions: \textit{non-vision-based} and \textit{vision-based} methods. Non-vision-based approaches\cite{li2024eyeecho,li2024auglasses,wu2021bioface,yao2025imuface} employ additional sensors such as EMG, EOG, or microphones, but these increase device weight, reduce comfort, and often suffer from limited accuracy. Vision-based methods\cite{olszewski2016high,patel2024fast,thies2016facevr,wei2019vr} instead use cameras to capture partial facial regions, yet they raise privacy concerns due to unintended body capture and face additional technical challenges, such as heterogeneous camera configurations across VR headsets and restricted access to raw data.

We tackle these limitations by exploiting the FACS\cite{ekman1978facial}-based blendshape signals provided by commercial VR headsets. These signals are derived from internal vision pipelines without exposing raw facial images, thereby reducing privacy concerns while still encoding core facial actions. Importantly, blendshapes require no additional sensors and are supported across most consumer VR headsets, ensuring broad applicability. Based on this observation, we present \textbf{OFERA} (Occluded Facial Expression to Realistic Avatar), a blendshape-driven framework that enables real-time expression control of photorealistic Gaussian head avatars for VR headset users. An overview of our system is shown in \cref{fig:teaser}.

Our contributions are threefold:
\begin{enumerate}
    \item \textbf{Expression Parameter Mapper (EPM):} an MLP-based model that maps headset-provided blendshape signals into an expression parameter space for controlling photorealistic Gaussian head avatars, enabling expressive control without raw camera access and addressing privacy and data access constraints.
    \item \textbf{Mapper-aware Data Adaptation:} a pipeline that aligns headset-specific blendshapes with the EPM distribution through Blendshape Distribution Alignment (BDA) and reduces train--test mismatch by integrating the EPM into avatar training via Mapper-integrated Avatar (MiA).
    \item \textbf{End-to-End Real-Time Gaussian Avatar Rendering Pipeline:} a complete pipeline that processes VR headset blendshape sensing, expression mapping, and Gaussian avatar updates to achieve real-time stereo rendering of photorealistic avatars in VR environments.
\end{enumerate}

We quantitatively validate that our MLP-based EPM outperforms prior mapping strategies in converting headset-provided blendshape signals into a controllable facial expression parameter space, achieving lower errors at both the parameter level and in vertex-based reconstructions. 
Beyond numerical evaluation, we conduct a user study in which participants assessed the expressiveness of their own avatars while wearing a VR headset. The results confirm that OFERA enables avatars to faithfully reproduce occluded facial expressions, and further demonstrate the effectiveness of our mapper-aware adaptation pipeline, where BDA reduces distribution mismatch and MiA ensures consistency during avatar training. Taken together, these findings show that OFERA not only advances the accuracy and robustness of expression mapping but also enhances user-perceived realism and expressiveness. By enabling photorealistic Gaussian avatars to reflect VR headset users' expressions in real-time, OFERA facilitates richer emotional exchange and more effective communication in immersive VR environments.

\section{Related Work}

\subsection{Avatars in VR Communication}
Avatars play a central role in enabling telepresence in VR environments, where the sense of being together depends on how faithfully avatars convey users' appearance, behavior, and expressions. Early research highlighted how avatars influence social dynamics and communication: Bailenson and Yee\cite{bailenson2004transformed} introduced the concept of Transformed Social Interaction, demonstrating that avatar-mediated behaviors can strongly affect social presence, while Slater et al.\cite{slater2010first} showed that embodying a virtual body can induce a strong sense of self-representation and agency.  

Subsequent studies examined how the realism and design of avatars shape user experience in communication. Bailenson et al.\cite{bailenson2006effect} investigated the effect of behavioral and form realism of real-time avatar faces, showing that more realistic facial rendering enhances perceived social presence. Baker et al.\cite{baker2021avatar} analyzed avatar-mediated communication in social VR and emphasized the importance of responsive avatars for interaction quality. More recent work\cite{combe2024exploring} explored the role of avatar heads and facial expressions in mixed reality collaboration, demonstrating that they significantly influence perceived social presence. Similarly, Lee et al.\cite{lee2023effects} reported that avatar visibility affects joint agency in collaborative tasks.  

Other studies further investigated how individual and contextual factors modulate avatar-mediated communication. Research has shown that gender differences influence the perception of avatar faces and interpersonal distance\cite{kang2024gender}, while collaboration context and personality traits also shape user experience in social VR\cite{kang2025collaboration}. In augmented and mixed reality settings, avatar appearance, body part representations, and visual fidelity have been shown to affect social presence, including avatar appearance\cite{yoon2019effect}, hand models\cite{yoon2020evaluating}, avatar transparency\cite{yoon2023effects}, and emotion-based prioritized facial expressions\cite{kang2024influence}. Beyond head-only representations, full-body generation approaches such as RC-SMPL\cite{song2023rc} demonstrated in user studies that avatar realism influences embodiment and user experience in VR.  

Together, these studies highlight that avatar design critically impacts social presence and telepresence in VR and mixed reality communication. Building on these insights, we focus on enhancing avatar expressiveness by enabling realistic facial expression control for VR headset users through our proposed framework, OFERA.

\begin{figure*} [!ht]
    \centering  \includegraphics[width=0.85\textwidth]{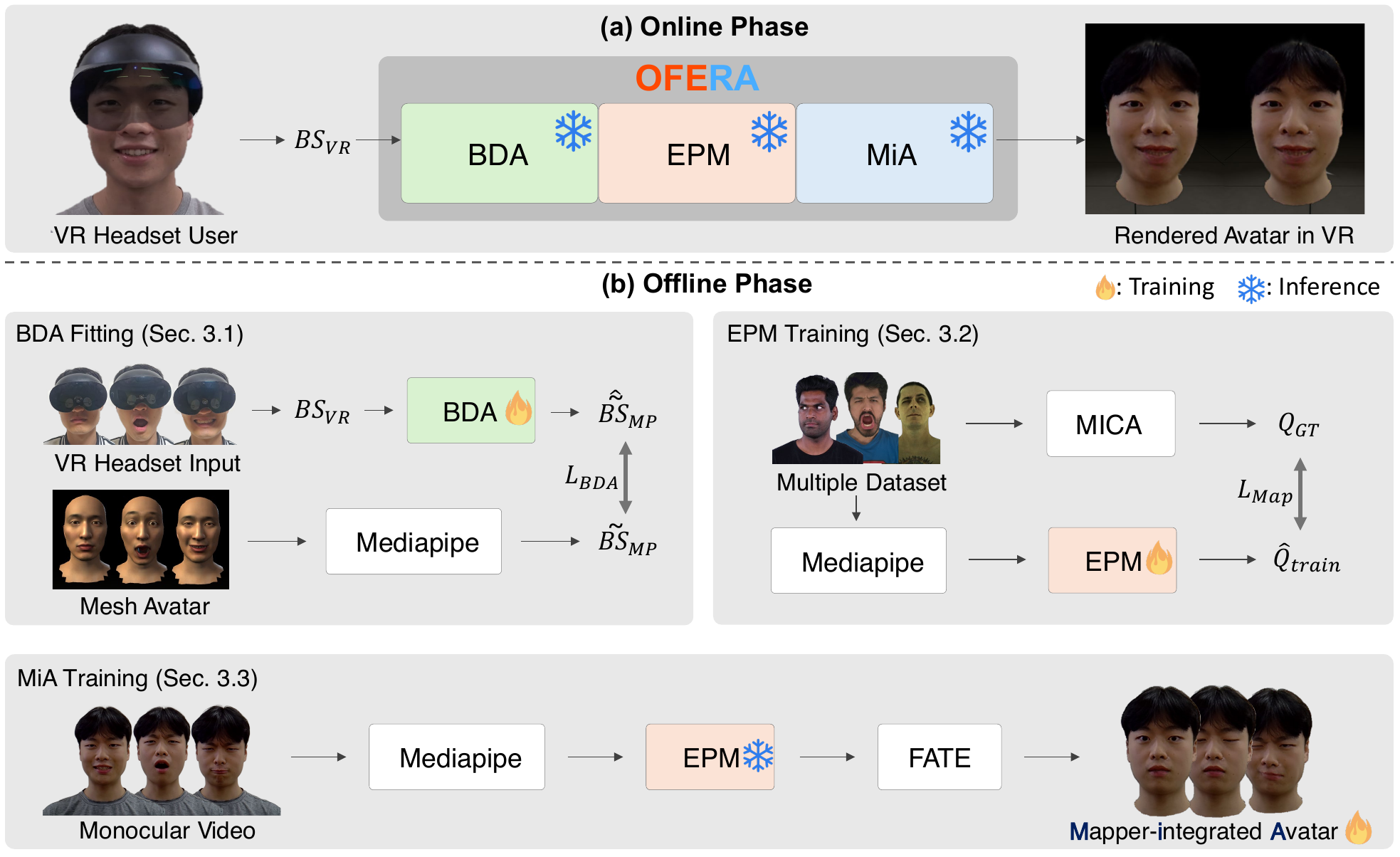}
  \caption{
    Overview of OFERA, which operates in online and offline phases. (a) In the online phase, headset blendshapes are processed through BDA, EPM, and MiA to drive a photorealistic 3D Gaussian avatar in real-time. (b) The offline phase involves fitting BDA (\cref{sec_bda}) and training EPM (\cref{sec_epm}) and MiA (\cref{sec_mia}) to ensure accurate expression mapping and consistent avatar rendering across headsets.
  }
  \label{fig:main}
\end{figure*}

\subsection{Head Avatar Reconstruction}
Realistic reconstruction of head avatars has long been a central research topic in computer vision and graphics, particularly for applications in virtual communication. The key goal is to create avatars that resemble users' identities while remaining animatable to capture facial expressions, which are essential for natural interaction.

The 3D Morphable Model (3DMM) by Blanz and Vetter\cite{blanz2023morphable} introduced a statistical framework for facial shape and texture modeling using PCA, enabling reconstruction from monocular images. FaceWarehouse\cite{cao2013facewarehouse} further expanded this direction with a large-scale dataset supporting identity–expression bilinear modeling. Li et al. later proposed FLAME\cite{li2017learning}, which disentangles identity, pose, and expression in a unified model with articulated jaw, neck, and eyes, and MICA\cite{zielonka2022towards} improved the encoding of identity with high-quality reconstruction. Other studies extended these models with unsupervised regression, hybrid losses, and affective expression modeling\cite{danvevcek2022emoca,deng2019accurate,feng2021learning, song2025fast,zhu2023facescape}. Mesh-based models thus provide compact and expressive control for avatar reconstruction and animation, and their low-dimensional parameters are often used to drive more detailed neural representations.

While mesh-based models provide efficient control, they are inherently limited in realism and fine-scale detail. Neural rendering has recently enabled photorealistic and animatable avatar reconstruction. NeRF\cite{mildenhall2021nerf} pioneered neural radiance fields for novel view synthesis, and subsequent works such as NerFace\cite{gafni2021dynamic}, AD-NeRF\cite{guo2021ad}, and Neural Head Avatars (NHA)\cite{grassal2022neural} extended this framework to dynamic and controllable head avatars. Additional studies further incorporated audio conditioning, parametric priors, or efficiency improvements\cite{duan2023bakedavatar,song2024tri,xu2023avatarmav,zielonka2023instant}. These methods achieved highly realistic avatars but remain impractical for real-time VR communication due to slow training and rendering. To address these limitations, 3D Gaussian Splatting (3DGS)\cite{kerbl20233d} emerged as an efficient alternative, supporting real-time rendering with high fidelity. Gaussian Head\cite{wang2025gaussianhead} demonstrated its applicability for head avatars, and GaussianAvatars\cite{qian2024gaussianavatars} generalized the approach across subjects. Many other extensions have advanced expression control, relightability, and efficiency\cite{chen2024monogaussianavatar,giebenhain2024npga,he2024emotalk3d,li2024uravatar,li2025rgbavatar,ma20243d,saito2024relightable,song2024toward,wang2025mega,zhang2025fate}, showing that 3DGS can represent dynamic heads with fine detail while supporting interactive frame rates, making them well-suited for VR communication scenarios.

In our work, we adopt FATE\cite{zhang2025fate} as the backbone model. FATE achieves state-of-the-art reconstruction quality and includes a full-head completion module that synthesizes plausible side and back views from monocular input, an essential property for VR avatars. We further integrate our Expression Parameter Mapper into the training process to enable consistent expression control.

\subsection{Facial Expression Capture with VR Headsets}
When users wear a VR headset, much of the face is occluded, making expression capture challenging. Prior work has explored both non-vision-based and vision-based methods.

Non-vision-based methods employ additional sensors to capture signals correlated with muscle activity or skin deformation. For example, AUGlasses\cite{li2024auglasses} and IMUFace\cite{yao2025imuface} use inertial sensing, BioFace-3D\cite{wu2021bioface} leverages EMG/EOG, and EyeEcho\cite{li2024eyeecho} applies ultrasonic sensing. Others explored RF antennas or body-mounted cameras\cite{chen2021neckface,chen2020c,kim2023pantoenna}. These approaches avoid direct imaging but often increase weight, require skin contact, and lack robustness in noisy environments. Vision-based methods integrate cameras into headsets to capture visible regions. Olszewski et al.\cite{olszewski2016high} regressed avatar parameters from headset cameras, FaceVR\cite{thies2016facevr} combined RGB-D and IR sensing for reenactment, and Wei et al.\cite{wei2019vr} and Patel et al.\cite{patel2024fast} advanced photorealistic facial animation. Other works further improved fidelity with codec avatars and appearance models\cite{bai2024universal,lombardi2018deep}. Beyond direct camera sensing, VOODOO XP\cite{tran2024voodoo} demonstrated that headset-provided blendshapes can be retargeted to a generic rigged mesh avatar and then used to drive photorealistic head reenactment for VR telepresence. Despite their robustness, these methods suffer from device-specific constraints and privacy issues since they require raw facial video streams.

In contrast, our system leverages blendshape parameters released by commercial VR headsets. Derived from internal vision pipelines, they preserve essential facial actions while avoiding privacy concerns. In particular, ARKit blendshapes are FACS\cite{ekman1978facial}-based and widely supported, making them an accessible and generalizable input for our framework, OFERA. Prior works have also attempted matrix-based or linear mappings from blendshapes to parametric model expressions\cite{liu2024emage,yan2024gaussian}, but these approaches are limited in expressiveness and generalization, which motivates our design of a learning-based Expression Parameter Mapper.

\section{Method}


Our framework, \textbf{OFERA} (Occluded Facial Expression to Realistic Avatar), provides an end-to-end pipeline that converts headset-derived blendshapes into photorealistic Gaussian head avatars in VR. As shown in \cref{fig:main}, headset blendshapes are first aligned to a canonical distribution by the Blendshape Distribution Alignment (BDA) module, then mapped to a continuous expression parameter space by the Expression Parameter Mapper (EPM), which is instantiated using the FLAME\cite{li2017learning} model in this work. The Mapper-integrated Avatar (MiA) ensures that avatars are trained and rendered consistently with the EPM outputs, and the resulting Gaussian updates are streamed to a stereo rendering pipeline in real-time. In this way, OFERA bridges device-dependent blendshape inputs with high-fidelity 3D Gaussian avatars, enabling natural and consistent expressions in VR communication.

\subsection{Blendshape Distribution Alignment (BDA)}
\label{sec_bda}

We use ARKit\footnote{Apple Developer, "ARFaceAnchor.BlendShapeLocation", \url{https://developer.apple.com/documentation/arkit/arfaceanchor/blendshapelocation}}-based blendshapes as the primary input to our framework. Since they are based on the Facial Action Coding System (FACS)\cite{ekman1978facial}, ARKit blendshapes can represent key facial expressions and are widely supported across various VR headsets.
Importantly, these blendshape signals are derived from internal headset pipelines and can be accessed without exposing raw facial images, making them well-suited for privacy-preserving facial expression capture under headset-based sensing.
However, ARKit blendshapes were originally designed for use with webcams or RGBD cameras, and the way different headsets provide ARKit-compatible coefficients is not standardized. For example, the Meta Quest Pro generates OVR\footnote{Meta Horizon, "Blendshape Visual Reference for Movement Extensions for OpenXR", \url{https://developers.meta.com/horizon/documentation/native/android/move-ref-blendshapes}} blendshapes from built-in sensors and then simply maps them to ARKit parameters with predefined scalar weights (e.g., $0.5$, $0.75$, or $1.0$). Such coarse remapping is sufficient for basic animation, but introduces significant discrepancies when these transformed ARKit values are used in learning-based pipelines such as our EPM, leading to increased prediction errors. As a result, the raw values obtained from VR headsets ($BS_{VR}$) are not directly compatible with the EPM trained on Mediapipe\cite{lugaresi2019mediapipe}-derived values.

To resolve this discrepancy, we introduce \textbf{Blendshape Distribution Alignment (BDA)}, a process that transforms VR headset blendshapes into a distribution suitable for EPM inference. Training BDA requires paired data $(BS_{VR}, \tilde{BS}_{MP})$, where $BS_{VR}$ are VR headset–driven ARKit blendshapes and $\tilde{BS}_{MP}$ are Mediapipe-style ARKit blendshapes.

Constructing such pairs directly is challenging, since a participant wearing a VR headset cannot provide unobstructed facial images for Mediapipe analysis. To overcome this, we built a pseudo paired dataset, which serves as a proxy for real paired samples; details are provided in the supplementary material. For each subject, we generated a mesh avatar using a commercial framework\footnote{Avaturn, "Avaturn", \url{https://avaturn.me}}, a tool that creates a personalized 3D mesh model from a few facial photographs. The generated avatar has a facial geometry and appearance closely resembling the subject (e.g., face shape and facial features) and comes with a pre-rigged set of blendshapes for expression control. While the subject performed various expressions with the headset on, the corresponding blendshapes ($BS_{VR}$) were recorded via the Meta Movement SDK\footnote{Meta Horizon, "Movement SDK for Unity - Overview", \url{https://developer.oculus.com/documentation/unity/move-overview}}. These blendshapes were then used to animate the mesh avatar, and Mediapipe was applied to the rendered frontal images to extract Mediapipe-style ARKit blendshapes, which we denote as $\tilde{BS}_{MP}$. This indirect strategy enabled us to approximate the required $(BS_{VR}, \tilde{BS}_{MP})$ pairs offline. The overall BDA fitting process is illustrated in \cref{fig:main}.

We note that the blendshape rig and amplitude calibration of the commercial mesh avatar are not guaranteed to be strictly consistent with those used by the VR headset. As a result, the constructed pseudo paired data may contain residual inter-model bias. Accordingly, BDA is not intended to establish a perfect semantic correspondence between headset-provided and Mediapipe-derived blendshapes, but rather to serve as a practical calibration step that mitigates large-scale distribution mismatch across heterogeneous input sources.

Using the constructed dataset, we fit a linear regression model to align $BS_{VR}$ with $\tilde{BS}_{MP}$:

\begin{equation}
    \label{eq:bda1}
    \hat{\tilde{BS}}_{MP} = W BS_{VR} + b
\end{equation}

where $W \in \mathbb{R}^{51 \times 51}$ and $b \in \mathbb{R}^{51}$ are regression parameters optimized by minimizing

\begin{equation}
    \label{eq:bda2}
    L_{BDA} = \frac{1}{N} \sum_{i=1}^{N} \| \tilde{BS}_{MP}^{(i)} - \hat{\tilde{BS}}_{MP}^{(i)} \|^2
\end{equation}

Since the regression model is lightweight and linear, BDA can be executed as a pre-processing module during inference with negligible overhead ($<1$ ms per frame). This ensures consistent expression recognition in VR environments and allows the same EPM to be reused across different input sources.

\subsection{Expression Parameter Mapper (EPM)}
\label{sec_epm}

\begin{figure} [t!]
\centering
  \includegraphics[width=0.99\columnwidth]{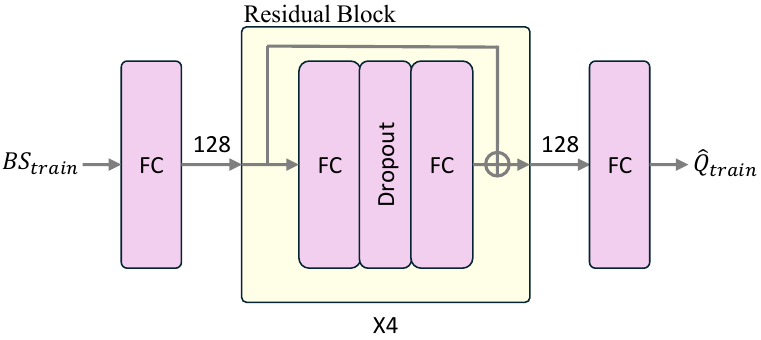}
  \caption{
    Architecture of the proposed EPM model. The network takes blendshape parameters ($BS_{train}$) as input and produces predicted target parameters ($\hat{Q}_{train}$) as output during training. It consists of an initial fully connected (FC) layer, followed by four residual blocks (each with two FC layers and an intermediate dropout layer), and a final FC layer. All FC layers are followed by batch normalization and ReLU activation, except for the last output FC layer. Each hidden FC layer has 128 units.
  }
  \label{fig:epm}
\end{figure}

To represent the facial expressions of VR headset users in a photorealistic Gaussian head avatar in real-time, we design an \textbf{Expression Parameter Mapper (EPM)} that maps headset-provided blendshape signals obtained in \cref{sec_bda} into a continuous expression parameter space suitable for controlling deformable Gaussian head avatars.
In this work, we instantiate this target expression parameter space using the FLAME\cite{li2017learning} model, and predict its expression parameters along with jaw and eye pose parameters, which are widely used in facial expression representation and compatible with existing avatar deformation pipelines.

Blendshapes represent human facial expressions as a combination of multiple parameters bounded within $[0,1]$. Different parameter combinations can yield visually similar expressions, which introduces ambiguity. Moreover, many parameters are strongly correlated (e.g., smiling involves simultaneous motion of lips, eyes, and cheeks), leading to complex dependencies. These properties make blendshape data highly non-linear. This implies that the transformation from ARKit blendshape coefficients to FLAME parameters cannot be sufficiently modeled as a simple linear regression problem. To address this, we adopt a mapping strategy based on a Multi-Layer Perceptron (MLP) architecture, as illustrated in \cref{fig:epm}. This design enables stable and accurate mapping of user expressions into the FLAME parameter space with low computational cost, allowing the system to operate without noticeable latency in VR headset environments.

For training the EPM, we use frontal face video datasets, as illustrated in \cref{fig:main}. For each frame, ARKit blendshape values $BS_{train} \in [0,1]^{51}$ are extracted using Mediapipe\cite{lugaresi2019mediapipe}, and ground-truth FLAME parameters $Q_{GT} \in \mathbb{R}^{68}$ are obtained using the MICA\cite{zielonka2022towards} model. The ground-truth parameters $Q_{GT}$ consist of FLAME expression parameters $e \in \mathbb{R}^{50}$ and the 6D rotations of the jaw, left eye, and right eye $(r_{jaw}, r^{L}_{eye}, r^{R}_{eye}) \in \mathbb{R}^{18}$. 

As shown in \cref{eq:epm1}, the EPM transforms $BS_{train}$ into predicted parameters $\hat{Q}_{train}$, and training is performed by minimizing the L1 loss $L_{Map}$ between $\hat{Q}_{train}$ and $Q_{GT}$ as defined in \cref{eq:epm2}:

\begin{equation}
  \label{eq:epm1}
  \hat{Q}_{train} = EPM(BS_{train})
\end{equation}
\begin{equation}
  \label{eq:epm2}
  L_{Map} = \lVert \hat{Q}_{train} - Q_{GT} \rVert_1 
\end{equation}

Since the model must generalize across diverse identities and a wide range of expression variations, we construct the training dataset by merging multiple sources and apply subject-wise sampling to mitigate distribution bias. This training process enables the EPM to effectively capture a broad distribution of facial expressions and generalize to unseen users.

\subsection{Mapper-integrated Avatar (MiA)}
\label{sec_mia}
Existing FLAME\cite{li2017learning}-based avatar training frameworks typically extract FLAME parameters from an input video, render the Gaussian avatar accordingly, and optimize the model by comparing the rendering with the ground-truth image. This approach inherently depends on the distribution of the extracted FLAME parameters, and is therefore strongly influenced by the parameter estimation method (e.g., MICA\cite{zielonka2022towards}).

In our system, we adopt the FATE\cite{zhang2025fate} model as the backbone for Gaussian avatar generation. FATE is particularly suitable for our setting, as it not only achieves high-quality novel view rendering but also incorporates a \textit{full-head completion} module that synthesizes plausible side and back views, which is crucial for VR scenarios. However, since FATE is optimized on the parameter distribution produced by MICA, any mismatch in parameter distribution during inference inevitably degrades rendering quality.

In practice, our framework relies on ARKit blendshape coefficients obtained from a frontal user video, denoted as $BS_{vid}$, which are converted into FLAME parameters by the EPM. The output distribution of the EPM, however, shows a discrepancy with that of MICA. Directly applying an avatar trained on MICA-based parameters to EPM outputs would amplify this mismatch, resulting in degraded expression rendering.

To address this issue, we propose the \textbf{Mapper-integrated Avatar (MiA)} strategy, where the EPM is directly integrated into the avatar training pipeline, as illustrated in \cref{fig:main}. Specifically, for each frame used during avatar training, ARKit blendshapes $BS_{vid}$ are passed through the EPM to obtain predicted FLAME parameters:

\begin{equation}
    \label{eq:mia1}
    \hat{Q}_{vid} = EPM(BS_{vid}) = (\hat{e}, \hat{r}_{jaw}, \hat{r}^{L}_{eye}, \hat{r}^{R}_{eye})
\end{equation}

These parameters are then used to deform the FLAME mesh following the formulation:

\begin{equation}
    \label{eq:mia2}
    T(\hat{Q}_{vid}) = LBS \Big( B_{P}(\Theta; \{\hat{r}_{jaw}, \hat{r}^{L}_{eye}, \hat{r}^{R}_{eye}\} + \Delta P) 
    \;+\; B_{E}(\Psi; \hat{e} + \Delta E) \Big)
\end{equation}

where $LBS$ denotes the linear blendshape skinning function, $B_P$ and $B_E$ denote the pose- and expression-dependent blendshapes of FLAME, respectively.

As a result, the loss terms originally defined in FATE are adapted such that the image reconstruction loss $L_{L1}$, perceptual loss $L_{vgg}$, and FLAME regularization $L_{flame}$ are all computed with respect to $T(\hat{Q}_{vid})$. We denote these modified terms as $L'_{L1}$, $L'_{vgg}$, and $L'_{flame}$. The complete training objective of MiA is thus given by:

\begin{equation}
    \label{eq:mia3}
    L = L'_{L1} + \lambda_1 L'_{vgg} + \lambda_2 L_{lap} + \lambda_3 L'_{flame} + \lambda_4 L_{scale}
\end{equation}

Here, $L_{lap}$ denotes Laplacian smoothing of mesh vertices, and $L_{scale}$ enforces constraints on Gaussian anisotropy. These regularization terms are independent of $\hat{Q}_{vid}$ and remain identical to those in the original FATE framework.

By aligning the avatar training process directly with the distribution of EPM outputs, the resulting Gaussian avatar can faithfully and consistently reflect users' expressions when driven by VR headset inputs at inference time.

\subsection{End-to-End Real-Time Gaussian Avatar System in VR}
\label{sec_render}
To enable end-to-end, real-time control of photorealistic Gaussian head avatars under headset-only sensing, we build a unified system that integrates lightweight communication between a Python server and a Unity client, along with expression mapping, avatar updates, and VR rendering.
During initialization, once the connection is established, a Gaussian avatar with a canonical expression is transmitted from the server to the client, and the Unity client performs stereo rendering of this idle avatar. In the execution stage, ARKit blendshape data $BS_{VR}$ are extracted from the VR headset via the Meta Movement SDK and sent to the server. Upon receiving the blendshapes, the server processes them through the BDA, EPM, and MiA modules to estimate the optimal FLAME\cite{li2017learning} parameters $\hat{Q}_{VR}$. These parameters are then used to update the Gaussian avatar, where each Gaussian\cite{kerbl20233d}’s position, rotation, and scale are modified according to the expression changes. Since opacity and spherical harmonics (SH) coefficients remain constant and are independent of expression changes, they are transmitted only once during initialization, and subsequent updates transmit only position, rotation, and scale. This optimization significantly reduces communication overhead and improves performance, enabling real-time interaction in VR environments. 

The Unity client supports dynamic updates of the Gaussian avatar by applying streamed expression-driven changes to Gaussian attributes and renders the avatar with the user’s expressions in real-time within the VR environment. The overall end-to-end latency from blendshape acquisition to avatar rendering is approximately $\sim$20ms, which is sufficient for interactive VR applications.
This system-level design enables practical deployment of Gaussian avatars in VR by bridging headset-derived expression signals and real-time avatar rendering, rather than introducing new avatar representations.

\begin{figure*}[!t]
\centering
 \includegraphics[width=\textwidth]{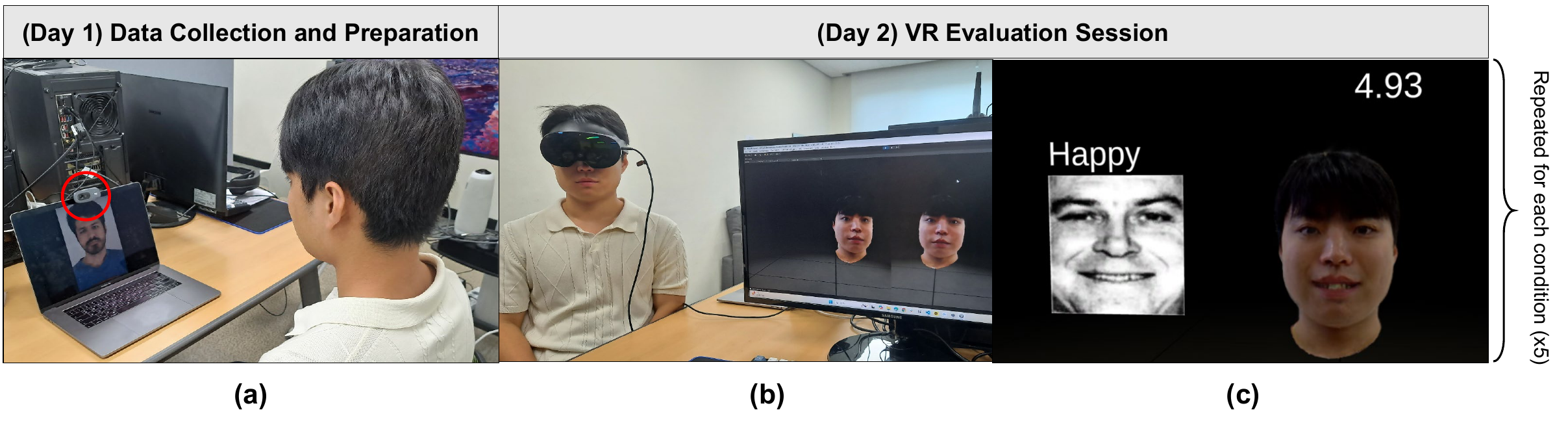}
\caption{
Overview of the two-day user study for facial data collection and VR-based evaluation (Day 1 and Day 2). (a) Facial expression recording for data collection and preparation; (b) VR setup for the evaluation session; and (c) an example study task in which participants controlled avatar facial expressions in VR using visual references. Avatar control and post-task questionnaires were repeated for each experimental condition (×5).
}
 \label{fig:setting}
\end{figure*}

\section{Experiment}

\subsection{Experimental Setup}
\textbf{Evaluation Design.\space\space}
Evaluating the proposed OFERA system requires both quantitative and qualitative assessments, as well as a user study. 
Since OFERA is designed to animate Gaussian head avatars in VR environments where users wear headsets, a ground-truth (GT) facial mesh captured under the same conditions is not accessible. 
Therefore, we distinguish two types of evaluation: 
(1) for the \textbf{EPM}, we perform quantitative and qualitative evaluation using front-facing facial images, from which ARKit blendshapes can be extracted via MediaPipe\cite{lugaresi2019mediapipe}; 
(2) for \textbf{BDA} and \textbf{MiA}, objective evaluation is infeasible, and we rely on a user study to validate their effectiveness in VR environments. 
Additionally, to evaluate the overall OFERA pipeline, we compare our system against baseline mapping approaches within the user study.

~\\
\textbf{Baselines.\space\space}
To validate the effectiveness of our MLP-based EPM module, we compare it with two existing approaches for mapping ARKit blendshapes to FLAME\cite{li2017learning} parameters.
Liu et al.\cite{liu2024emage} introduced a \textbf{matrix-based mapping} method, where a fixed $51\times103$ matrix transforms 51-dimensional ARKit blendshapes into 100 expression parameters and 3 jaw rotations of the FLAME model via direct multiplication. 
This approach is simple and efficient but limited by its inability to capture non-linear dependencies.
Yan et al.\cite{yan2024gaussian} proposed a \textbf{ridge regression-based linear mapping}, where ARKit blendshapes are linearly projected to FLAME parameters, including 100 expressions, 3 jaw rotations, and 6 eye rotations. 
While the ridge regularization improves stability, the linear model cannot represent the complex interactions among blendshape parameters.
These two mapping strategies serve as baselines for evaluating our EPM.

~\\
\textbf{Dataset.\space\space}
To train the EPM module across diverse distributions of facial identity and expression, we combine multiple datasets: INSTA\cite{zielonka2023instant}, NeRSemble\cite{kirschstein2023nersemble}, and Ava-256\cite{martinez2024codec}. 
From each dataset, we randomly select 10 subjects, resulting in 30 subjects in total. 
Each dataset is split into 8 subjects for training, 1 for validation, and 1 for testing, ensuring subject-disjoint partitions and preventing distributional bias across datasets.
We specifically use front-facing images of the test subjects for evaluation: 
ground-truth FLAME\cite{li2017learning} parameters are obtained using the MICA\cite{zielonka2022towards} tracker, and ARKit blendshapes are extracted from the same frames using MediaPipe\cite{lugaresi2019mediapipe}. 
The extracted blendshapes are then mapped into FLAME parameters using each baseline and our EPM, enabling direct comparison against the ground-truth FLAME results. 
The detailed dataset statistics are provided in the supplementary material. 

~\\
\textbf{Metrics.\space\space}
We evaluate the accuracy of expression mapping by computing the Root Mean Square Error (RMSE) between the predicted and ground-truth FLAME\cite{li2017learning} parameters.
In addition, since all FLAME meshes share consistent topology and coordinate systems, we further compute the vertex-wise position error as RMSE.
For quantitative evaluation, eye pose parameters were excluded from all methods, as the matrix-based mapping baseline does not predict eye-related parameters. Accordingly, errors were computed using only expression and jaw parameters to ensure a fair comparison across methods.
To facilitate qualitative analysis, we visualize the vertex error distribution as heatmaps, allowing us to inspect localized differences in facial geometry.
This evaluation protocol enables us to assess not only the parameter-level accuracy of expression mapping but also the geometric fidelity of reconstructed meshes.

\subsection{User Study}


We conducted a subjective user evaluation to assess the effectiveness of the proposed system in reflecting users' actual facial expressions on a virtual avatar. Specifically, the study examined the extent to which avatar facial expressions were realistically conveyed in VR environments and whether facial parts occluded by a headset could also be accurately represented. 
To this end, five expression mapping methods were set as experimental conditions to compare with our system, including two baselines and two ablations, apart from ours.
For each condition, we generated a realistic avatar model based on the participant’s real face and applied each mapping method. In all conditions, participants wore a headset and evaluated the facial expressions of their self-representative avatars in a VR environment.
All user study experiments were conducted using a Meta Quest Pro headset.

~\\
\textbf{Study Design and Task.\space\space}
The user experiment was designed with expression mapping method (\textit{method}) as the experimental factor, and five conditions were derived: (1) Matrix-based mapping (\textit{matrix}), (2) Linear-based mapping (\textit{linear}), (3) Ours without BDA (\textit{w/o BDA}), (4) Ours without MiA (\textit{w/o MiA}), and (5) OFERA (\textit{Ours}). Among them, (1) \textit{matrix} and (2) \textit{linear} were set as baseline conditions for direct comparison with our system, while (3) \textit{w/o BDA} and (4) \textit{w/o MiA} were configured as ablation model conditions. 
The \textit{method} factor was treated as a within-subject factor, such that each participant experienced all five conditions. To reduce potential order effects, the presentation order of the five conditions was counterbalanced using a balanced Latin Square method.

For the evaluation, participants were asked to perform a task while wearing a VR headset: They performed facial expressions by imitating given reference materials (images or short video clips). The avatar’s facial expressions changed in real-time according to the participant’s actual expressions, and participants observed and evaluated their self-representative avatar. Each avatar was generated in advance by scanning the participant’s face so that the avatar closely resembled their own appearance. To allow participants to concentrate solely on facial movements and expression, only a head avatar was used. 
The reference stimuli consisted of emotion-related images and short video clips, each presented for a fixed duration to allow consistent and sufficient observation. 
The overall task sequence consisted of two parts: (1) trying to imitate universal facial expressions raised by Paul Ekman\footnote{Paul Ekman Group, “Universal Facial Expressions”, \url{https://www.paulekman.com/resources/universal-facial-expressions/}}, and (2) trying to imitate more detailed partial expressions. The latter was designed to examine facial actions in specific parts of the face, informed by the Facial Action Coding System (FACS)\cite{ekman1978facial}, which was developed to categorize facial muscle actions associated with emotions. Among FACS, we selected emotion-related action units and also included commonly observed expressions---such as eye movements (e.g., eyes, brow, lid), cheek, nose, lips, and jaw movements---to better capture natural facial behaviors. 
All participants experienced the same sequence in the order of (1) universal expressions followed by (2) partial expressions, and were thus exposed to a wide variety of expressions for evaluation. 

\begin{figure*}[!t]
\centering
 \includegraphics[width=0.85\textwidth]{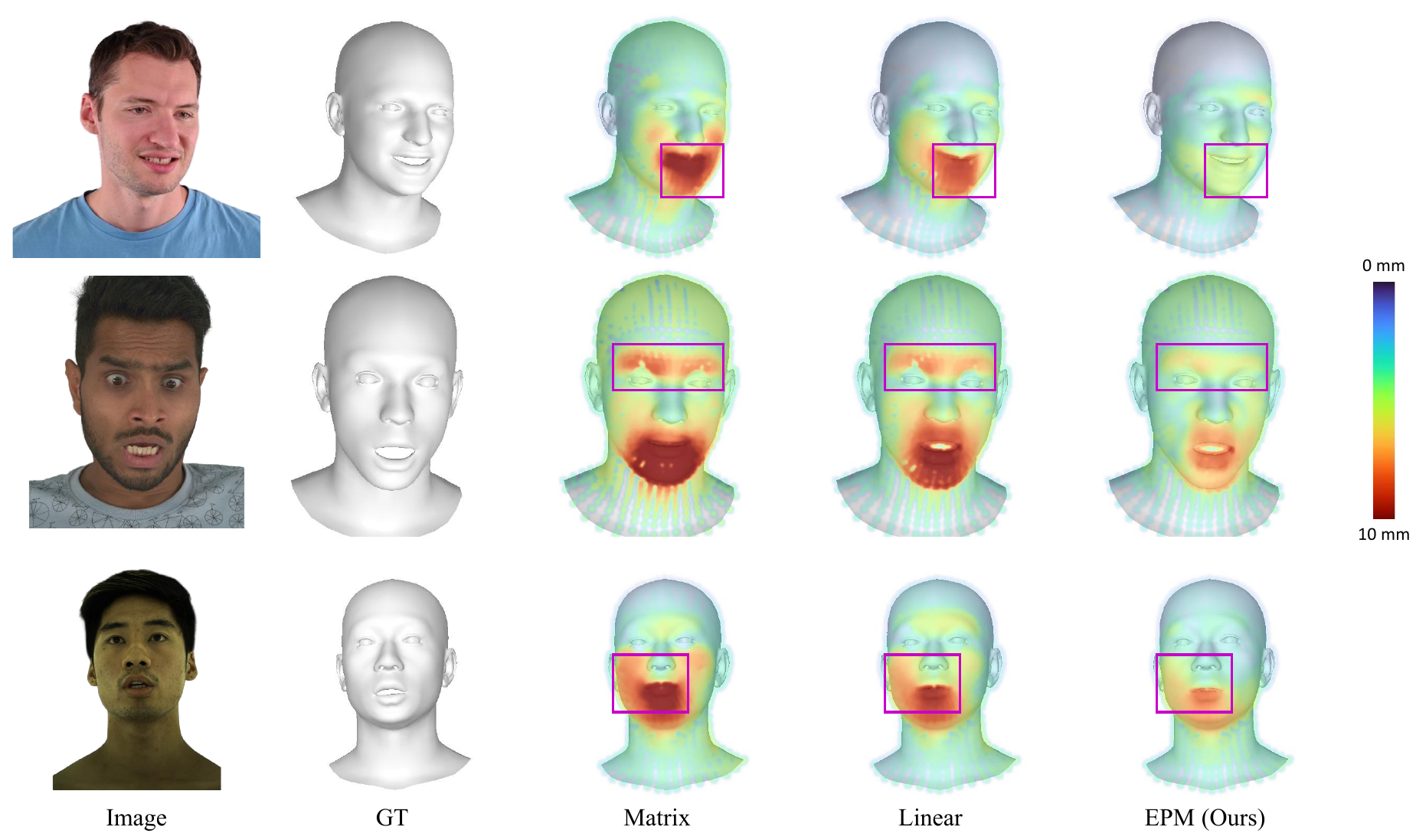}
\caption{
Qualitative comparison of vertex-wise reconstruction errors visualized as heatmaps. 
Matrix-based and linear mapping baselines show large errors around expressive regions such as the mouth and eyes, 
whereas our EPM significantly reduces these localized errors, leading to more faithful reproduction of facial expressions.
}
 \label{fig:qual}
\end{figure*}

~\\
\textbf{Dependent Variables.\space\space}
To evaluate how well a participant’s own expressions were reflected in their avatar and how controllable the expressions were while wearing a VR headset, we employed Sense of Embodiment (SoE) as the main dependent variable, selecting two widely used measurements from previous studies investigating avatar embodiment and facial expressions in virtual environments\cite{weidner2023systematic, gonzalez2020using, kullmann2025coverage}. 
First, we adopted the Virtual Embodiment Questionnaire (VEQ) proposed by Roth and Latoschik\cite{roth2020construction}. Among its subscales, we used Virtual Body Ownership (VBO), measuring the extent to which users perceived the virtual body as their own, and Agency (AG), evaluating the sense of control over the virtual body. 
Second, we employed the extended VEQ (VEQ+) measurement\cite{fiedler2023embodiment} to capture Self-Identification (SI), which reflects identifying the virtual representation as oneself. We assessed two SI-related subscales: Self-Attribution (SA) and Self-Similarity (SS), both indicating how strongly the avatar is perceived to represent the user in terms of personalization and similarity. VBO and AG from VEQ, and SA and SS from VEQ+ comprised a total of 16 items (4 items each).  

Next, we used the Virtual Human Plausibility Questionnaire (VHPQ) to assess the naturalness and coherence of avatars in the VR scene\cite{mal2022virtual}. Specifically, we included the Appearance and Behavior Plausibility (ABP) subscale (6 items) to evaluate the plausibility of the avatar’s visual appearance and motion behavior. Finally, the Facial Animation Realism (Real) measurement was adopted based on prior studies\cite{kang2024influence, kang2024gender, fraser2022expressiveness}, consisting of 4 items that assess the realism and naturalness of facial expressions and movements. All questionnaire items were rated on a 7-point Likert scale, and subscales with low relevance or items unmeasurable due to the study setup and purpose were excluded.
At the end of the experiment, we conducted open-ended interviews to gather subjective feedback. The questions focused on which of the five avatar models participants found most similar to their own face and expressions, most expressive, most natural, and realistic. Participants were also asked which avatar they would prefer to use as their own representation and which version they considered the best overall.

~\\
\textbf{Procedure.\space\space} 
Prior to the experiment, approval for the study content and procedure was obtained from the Institutional Review Board (IRB). The study consisted of two phases, which were conducted on separate days with at least a one-day interval between them.
In the first phase, participants visited for the collection of facial expression data required for avatar reconstruction. During this session, videos of participants performing various facial expressions were recorded. After that, they were given a detailed explanation about the study's purpose and procedure, followed by informed consent and a demographic questionnaire.

In the second phase, participants completed the main task of evaluating avatar facial expressions and movements. 
At the beginning of this session, the experimenter briefly explained the procedure, and participants were seated and wore a VR headset. Following the on-screen instructions, they tried to make the same facial expressions as those shown in reference images or videos while observing their avatar in VR. An overall illustration of the experiment is shown in \cref{fig:setting}. 

The five facial expression mapping methods were presented in a counterbalanced order, and thus the same task was repeated for all five conditions. After each method condition, participants completed a post-task questionnaire based on their observations. Once all conditions were completed, they removed the VR headset and participated in a post-experiment interview, where they shared their overall feedback on each method.

~\\
\textbf{Participants.\space\space} 
We recruited 20 participants (10 male, 10 female) through the university’s community and website. The participants’ mean age was 27.8 years ($SD =$ 3.17): 27.2 ($SD =$ 2.94) for males and 28.4 ($SD =$ 2.44) for females.
As the study involved observing avatar facial expressions in VR, we also asked about participants’ prior experiences with related technologies: Regarding avatar-mediated technologies or applications, 16 reported less than four prior uses (80\%), among whom five had no experience at all (25\%), while four participants reported more than ten prior uses (20\%). For AR/VR-related experiences (e.g., wearable devices, applications), four participants reported less than four prior uses, whereas 16 had more than five experiences, including 11 who had used them more than ten times (55\%).

\section{Results}

\subsection{Expression Mapping Evaluation}
\label{sec:expr-mapping-eval}

\Cref{table:quan} summarizes the quantitative performance of different expression mapping methods. 
Among all methods, the matrix-based mapping approach yields the highest errors. 
The ridge regression-based linear mapping improves performance over the matrix-based approach. 
In contrast, our proposed EPM achieves the lowest errors in both parameter space (0.505) and vertex space (1.593), outperforming both baselines.
Beyond these quantitative metrics, \cref{fig:qual} visualizes the vertex-wise error distributions across different methods. 
Consistent with the quantitative results, the matrix-based and linear mappings exhibit substantial deviations around highly expressive regions such as the mouth and eyes, whereas our EPM shows reduced errors in these regions.
This trend is particularly noticeable in expressions involving large facial deformations.

\begin{table}[!t]
\caption{
Quantitative comparison of expression mapping methods. 
Our EPM consistently achieves the lowest parameter- and vertex-level errors, 
demonstrating the benefit of modeling non-linear dependencies between ARKit blendshapes and FLAME parameters.
}
\label{table:quan}
\centering
\renewcommand{\arraystretch}{1.6} 
\begin{tabular}{lcccc}
\hline
Mapping Method & \textbf{Param Error} $\downarrow$ & \textbf{Vertex Error (mm)} $\downarrow$ \\
\hline
Matrix & 1.038 & 2.826 \\
Linear & 0.573 & 1.848 \\
\textbf{EPM (Ours)} & \textbf{0.505} & \textbf{1.593} \\
\hline
\end{tabular}
\end{table}


\begin{figure*}[!ht]
\centering
\includegraphics[width=\textwidth]{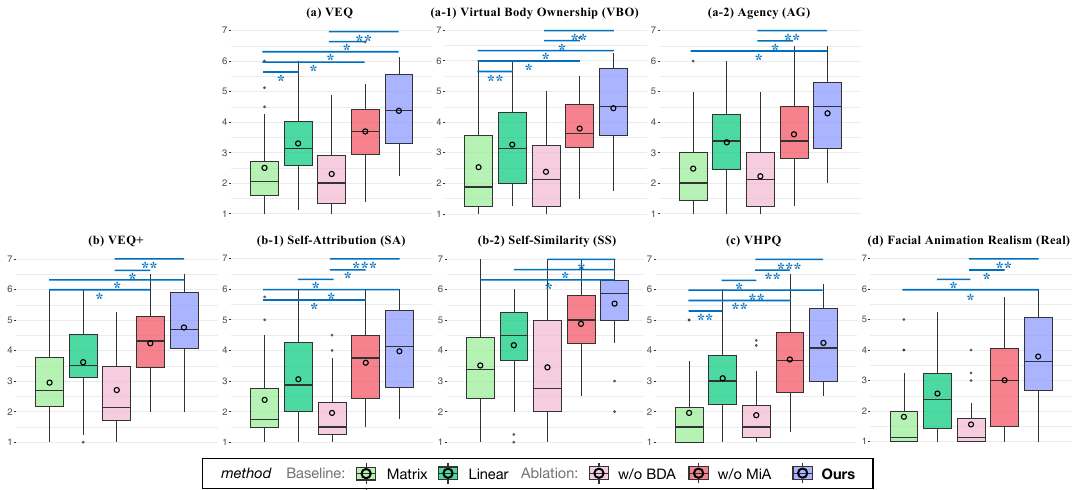}
\caption{Results for (a)--(a-2) Virtual Embodiment Questionnaire and subscales; (b)--(b-2) VEQ+ and the subscales associated with Self-Identification; (c) Virtual Human Plausibility Questionnaire; (d) Facial Animation Realism.}
 \label{fig:graph}
\end{figure*}

\subsection{User Study Analysis}
Because the user study involved subjective measures across five within-subject conditions, we used a Friedman test for non-parametric analysis ($\alpha =$ .05). Post-hoc pairwise comparisons were conducted using a Wilcoxon Signed Rank test with Bonferroni correction. Effect sizes are reported using Kendall’s W, with values of approximately .10, .30, and .50 indicating small, moderate, and large effects, respectively. The reliability of the Likert-scale items was verified using Cronbach’s alpha. The statistical results are illustrated in \cref{fig:graph}.

\textbf{VEQ: } 
The internal consistency of VEQ and the two subscales (VBO and AG) all showed an acceptable range ($\alpha_{VEQ} =$ .951; $\alpha_{VBO} =$ .901; $\alpha_{AG} =$ .947). 
A significant main effect of \textit{method} on VEQ scores was found ($\chi^2(4) =$ 33.841, $p<$ .001, $W =$ .423). 
Pairwise comparisons revealed significant differences between \textit{Ours} and \textit{matrix} ($p = .011$), and \textit{Ours} and \textit{w/o BDA} ($p=$ .002). Additional differences revealed in the pairs of \textit{matrix} and \textit{linear} ($p=$ .046), \textit{matrix} and \textit{w/o MiA} ($p=$ .025), and \textit{w/o BDA} and \textit{w/o MiA} ($p=$ .022). 
For VBO, the main effect of \textit{method} was also significant ($\chi^2(4) =$ 27.969, $p<$ .001, $W =$ .350). Post-hoc revealed significant contrasts in the pairs of \textit{Ours} and \textit{matrix} ($p=$ .018), and \textit{Ours} and \textit{w/o BDA} ($p=$ .003). Additional pairwise differences were observed for \textit{matrix} and \textit{linear} ($p=$ .009), \textit{matrix} and \textit{w/o MiA} ($p=$ .033), and \textit{w/o BDA} and \textit{w/o MiA} ($p=$ .023) pairs.   
Regarding the subscale AG, a significant effect of method was also observed ($\chi^2(4) =$ 30.175, $p<$ .001, $W =$ .377). During the post-hoc analysis, significant differences were found between \textit{Ours} and \textit{matrix} ($p=$ .022), and \textit{Ours} and \textit{w/o BDA} ($p=$ .003), along with a difference for \textit{w/o BDA} and \textit{w/o MiA} ($p=$ .031). All other pairs did not significantly differ on VEQ, VBO and AG scores. 


~\\
\textbf{VEQ+: } 
The VEQ+ scale and its subscales demonstrated high reliability  ($\alpha_{VEQ+} = .959$; $\alpha_{SA} = .932$; $\alpha_{SS} = .975$). The Friedman test indicated a significant effect for method on VEQ+ ($\chi^2(4) =$ 26.588, $p<$ .001, $W =$ .332). In post-hoc analysis, significant differences were found between \textit{Ours} and \textit{matrix} ($p=$ .010), and \textit{Ours} and \textit{w/o BDA} ($p=$ .003). Further contrasts appeared for \textit{matrix} and \textit{w/o MiA} ($p=$ .028), and \textit{w/o BDA} and \textit{w/o MiA} ($p=$ .022). 
Turning to SA, the main effect of method was also significant ($\chi^2(4) =$ 39.947, $p<$ .001, $W =$ .499). Post-hoc tests showed that \textit{Ours} and \textit{matrix} ($p=$ .022), and \textit{Ours} and \textit{w/o BDA} ($p<$ .001) conditions were significantly different. Other differences were found in the following pairs: \textit{matrix} and \textit{w/o MiA} ($p=$ .019), \textit{linear} and \textit{w/o BDA} ($p=$ .015), and \textit{w/o BDA} and \textit{w/o MiA} ($p=$ .012).  
For SS, a significant main effect was found for \textit{method} ($\chi^2(4) =$ 22.827, $p<$ .001, $W =$ .285). Post-hoc analysis only revealed significant differences between \textit{Ours} and \textit{matrix} ($p=$ .015), \textit{linear} ($p=$ .041), and \textit{w/o BDA} ($p=$ .022). No other significant differences were observed for VEQ+, SA, and SS. 

~\\
\textbf{VHPQ: } 
We utilized a subscale ABP (Appearance and Behavior Plausibility) of VHPQ, and its internal consistency was satisfied with the accepted level ($\alpha_{VHPQ} =$ .945). A significant main effect of \textit{method} was found ($\chi^2(4) =$ 45.522, $p<$ .001, $W =$ .569), and post-hoc tests showed significant differences in the pairs of \textit{Ours} and \textit{matrix} ($p=$ .006), \textit{Ours} and \textit{w/o BDA}  ($p<$ .001). Significant contrasts were also identified for \textit{matrix} and \textit{linear} ($p=$ .008), \textit{matrix} and \textit{w/o MiA} ($p=$ .005), \textit{linear} and \textit{w/o BDA} ($p=$ .016), and \textit{w/o BDA} and \textit{w/o MiA} ($p=$ .009). All other pairs were not significantly different.

~\\
\textbf{Facial Animation Realism: }
Lastly, the Real scale demonstrated excellent internal consistency ($\alpha_{Real} =$ .963). We found a significant main effect of \textit{method} ($\chi^2(4) =$ 37.130, $p<$ .001, $W =$ .464). Pairwise comparisons revealed significant differences between \textit{Ours} and \textit{matrix} ($p=$ .018), and \textit{Ours} and \textit{w/o BDA} ($p=$ .001). Other pairs that showed significant differences were \textit{linear} and \textit{w/o BDA} ($p=$ .017), and \textit{w/o BDA} and \textit{w/o MiA} ($p=$ .028). No other pairs showed significant differences on Real.

\section{Discussion}


Our experimental results indicate that OFERA consistently outperforms the baselines (\textit{matrix}, \textit{linear}) and ablation models (\textit{w/o BDA}, \textit{w/o MiA}) in both objective expression mapping accuracy and subjective user assessments.
In the quantitative and qualitative evaluations of expression mapping, OFERA achieved lower parameter and vertex errors than the \textit{matrix}- and \textit{linear}-based baselines, with particularly notable improvements in highly dynamic facial regions such as the eyebrows and mouth.
These results indicate that transforming blendshape signals into a controllable expression parameter space benefits from non-linear modeling, and support the design choice of adopting an MLP-based Expression Parameter Mapper (EPM).
In particular, the reduced vertex-wise reconstruction errors and clearer geometric structures observed in mesh-level visualizations suggest that the proposed non-linear mapping better captures the underlying geometric and structural characteristics of facial expressions than linear alternatives.
Building on these objective improvements in expression mapping, we further examine how such gains translate into perceived realism, controllability, and embodiment in immersive VR environments through a user study.

These advantages were further reflected in the user study.
In the statistical analysis, our method achieved significantly higher scores in virtual embodiment, self-identification, plausibility, and facial animation realism compared to \textit{matrix}-based mapping and \textit{w/o BDA} ablation conditions. Compared to the more competitive models (\textit{linear} and \textit{w/o MiA}), \textit{Ours} reduced distortions more effectively than \textit{linear} and preserved richer expressiveness than \textit{w/o MiA}, leading to higher overall embodiment and realism. Since \textit{w/o MiA} represents an ablation setting, the differences between \textit{w/o MiA} and \textit{Ours} were expected to be smaller, yet \textit{Ours} still showed an advantage in expressiveness and controllability. 

These trends were confirmed by participant feedback. \textit{Ours} was described as \textit{``most natural and human-like''}(P2), \textit{``best followed my face and expressions''}(P8), and \textit{``accurately reflected even subtle muscle movements''}(P5). In contrast, \textit{linear} was noted for detail but still \textit{``heavily distorted''}(P11), while the \textit{w/o MiA} model was stable but \textit{``less expressive''}(P13). As a result, these observations indicate that \textit{Ours} achieves better performance in terms of stability and expressiveness. 
Finally, our system also demonstrated practical usability in VR, even under headset occlusion. Unlike baseline conditions that failed to reproduce eye movements (\textit{``the eyes did not move at all''}, P18), \textit{Ours} could reproduce such facial dynamics. 
Overall, these findings demonstrate that \textit{Ours} provides a solid foundation for realistic, controllable virtual avatar expressions, while also delivering greater embodiment and self-identification to users than alternative mapping methods. 

Beyond mapping accuracy and perceptual realism, our findings highlight the importance of system-level integration for practical avatar deployment in VR.
Rather than introducing a new avatar representation, OFERA demonstrates that combining headset-available sensing, calibrated parameter mapping, and lightweight real-time communication can effectively bridge the gap between occluded user input and expressive photorealistic avatars.
This suggests that realistic avatar control in VR can be achieved through careful coordination across sensing, representation, and rendering stages, even under strict latency and access constraints.
Such a system-oriented perspective complements recent advances in avatar modeling and is particularly relevant for deployable VR telepresence applications.

\subsection{Limitations and Future Work}

The proposed system involves a stack of constraints arising from multiple stages of representation and parameter transformation, which can limit the expressiveness of the final avatar despite explicit mitigation efforts.
In particular, OFERA relies on headset-provided blendshape signals as the sole input modality, inherently constraining expressiveness to the capacity of the blendshape space and limiting the accurate reproduction of subtle facial motions such as micro-expressions.
These constraints are further influenced by differences in blendshape definitions and amplitude calibration across VR headsets, as well as discrepancies between the Meta Quest Pro blendshape pipeline and the commercial mesh avatar used for pseudo paired data construction.
Although Blendshape Distribution Alignment (BDA) and Mapper-integrated Avatar (MiA) are introduced to reduce distribution mismatch and training--inference inconsistency across stages, residual constraints may still propagate through the pipeline.
Moreover, the fidelity of facial deformations ultimately depends on the representation capacity of the underlying Gaussian avatar backbone model\cite{zhang2025fate}, and certain expression patterns may not be fully captured if the backbone is insufficiently trained.
At the system level, to ensure real-time performance, OFERA fixes spherical harmonics and opacity during runtime, which limits support for dynamic relighting or appearance variations.
Finally, our user study focused on observing overall user perceptions of the proposed system and therefore did not explicitly investigate other factors related to avatar representation, such as age, individual facial characteristics (e.g., face shape and expression style), or environmental lighting.

Future work will explore richer or more unified expression representations that reduce stacked pipeline constraints while preserving real-time performance, as well as more efficient system designs that enable dynamic appearance attributes.
We also plan to extend the framework to support cross-headset calibration and evaluate robustness across heterogeneous devices, along with broader user studies conducted under more diverse participant profiles and usage conditions.

\section{Conclusion}

We propose \textbf{OFERA} (Occluded Facial Expression to Realistic Avatar), a real-time system that reconstructs occluded facial expressions of VR headset users as photorealistic 3D Gaussian avatars. OFERA is composed of three modules: (i) the \textit{Blendshape Distribution Alignment} (BDA), which adapts headset-specific blendshapes to a canonical input space suitable for inference; (ii) the \textit{Expression Parameter Mapper} (EPM), which maps the aligned blendshape signals into an expression parameter space for controlling Gaussian head avatars; and (iii) the \textit{Mapper-integrated Avatar} (MiA), which ensures that the Gaussian avatar is trained to follow the output distribution of the EPM.
By integrating these modules, OFERA establishes an end-to-end pipeline that takes blendshape data captured from a VR headset and renders a Gaussian avatar with realistic expressions in real-time within the VR environment. Our experiments demonstrated that the proposed EPM outperforms baseline models in both quantitative and qualitative evaluations, and user studies further validated the effectiveness of the BDA and MiA modules. This confirms that OFERA enables both realistic appearance and natural real-time expression control, which are critical for immersive telepresence in VR.

\acknowledgments{%
This work was supported by the National Research Council of Science \& Technology (NST) grant by the Korea government (MSIT) (No. CRC21015);
This work was supported by the Institute of Information \& Communications Technology Planning \& Evaluation(IITP) grant funded by the Korea government(MSIT) (No. RS-2025-25441313, Professional AI Talent Development Program for Multimodal AI Agents);
This research was supported by the MSIT(Ministry of Science and ICT), Korea, under the Graduate School of Metaverse Convergence support program(IITP-2026-RS-2022-00156435) supervised by the IITP(Institute for Information \& Communications Technology Planning \& Evaluation).
}

\bibliographystyle{abbrv-doi-hyperref}

\bibliography{template}

\end{document}